\documentclass[10pt,letterpaper]{article}
\usepackage[top=0.85in,left=1.75in,footskip=0.75in]{geometry}

\usepackage{amsmath,amssymb}

\usepackage{changepage}

\usepackage[utf8x]{inputenc}

\usepackage{textcomp,marvosym}

\usepackage{cite}

\usepackage{nameref,hyperref}

\usepackage[right]{lineno}

\usepackage{microtype}
\DisableLigatures[f]{encoding = *, family = * }

\usepackage[table]{xcolor}

\usepackage{array}

\newcolumntype{+}{!{\vrule width 2pt}}

\newlength\savedwidth

\usepackage[aboveskip=1pt,labelfont=bf,labelsep=period,justification=raggedright,singlelinecheck=off]{caption}

\makeatletter
\renewcommand{\@biblabel}[1]{\quad#1.}
\makeatother

\usepackage{lastpage,fancyhdr,graphicx}
\usepackage{epstopdf}
\fancyheadoffset[L]{2.25in}

\usepackage{multirow}

\begin{document}
\vspace*{0.2in}

\begin{flushleft}
{\Large
\textbf\newline{ShinyCOPASI: a web-based exploratory interface for COPASI models.} 
}
\newline
\\
Abhishekh Gupta,
Pedro Mendes
\\
\bigskip
Center for Cell Analysis and Modeling and Department of Cell Biology,\\
University of Connecticut School of Medicine, \\
263 Farmington Av., Farmington, CT 06030-6033, USA \\

%
%

Correspondence: pmendes@uchc.edu

\end{flushleft}
\section*{Abstract}
COPASI is a popular application for simulation and analysis of biochemical networks and their dynamics. While this software is widely used, it works as a standalone application and until now it was not possible for users to interact with its models through the web. We built ShinyCOPASI, a web-based application that allows COPASI models to be explored through a web browser. ShinyCOPASI was written in R with the CoRC package, which provides a high-level R API for COPASI, and the Shiny package to expose it as a web application. The web view provided by ShinyCOPASI follows a similar interface to the standalone COPASI and allows users to explore the details of a model, as well as running a subset of the tasks available in COPASI from within a browser. A generic version allows users to load model files from their computer, while another one pre-loads a specific model from the server and may be useful to provide web access to published models. The application is available at: \href{http://shiny.copasi.org/}{http://shiny.copasi.org/}; and the source code is at: \href{https://github.com/copasi/shinyCOPASI}{https://github.com/copasi/shinyCOPASI}.


\section*{Introduction}
COPASI is a stand-alone software application for simulation and analysis of biochemical networks and their dynamics that includes powerful algorithms yet is based on a graphical user interface that makes it easy to use without expertise in programming or mathematics~\cite{Hoops_2006,Bergmann_2017}. It supports models in the SBML standard~\cite{Hucka2003} and can simulate their behavior using ordinary differential equations, stochastic differential equations, or Gillespie's stochastic simulation algorithm.  

COPASI has been used in published research in many fields of biology and chemistry~\cite{Copasi_research} and yet there is no easy way to expose those models to readers online in a functional way. Currently readers need to install COPASI in their computers and download supplementary files to be able to examine models published as supplementary data. A light-weight web interface to allow exploration of COPASI models is highly desirable. Such a tool would allow readers of research articles to quickly interact with  published models, including running some simulations or analyses (those that are not time intensive). It would additionally facilitate the dissemination of models and their associated results, and promote reproducibility~\cite{Mendes2018}.

Here, we present ShinyCOPASI, a web-based application that provides online access to functional COPASI models. It exists in two variants, a {\it generic} interface at \href{http://shiny.copasi.org/}{http://shiny.copasi.org/} that allows users to load and interact with any COPASI or SBML file, and a {\it specialized} version to be mounted on any web site and which provides access to a predetermined model file. The latter is designed to allow exposing specific model files on web sites, for example those that are included in research publications. This follows the approach pioneered by JWS-online~\cite{Olivier2004} but specifically for COPASI models and without requiring any recoding.

\section*{Implementation}\label{Implementation}

ShinyCOPASI is implemented in the R programming environment~\cite{R-base} and depends on several packages that are automatically installed and loaded when it is executed. The ability of ShinyCOPASI to interact with users through a web interface is provided by the {\it Shiny} package, which is a web application framework for R \cite{R-shiny}. It allows building interactive web applications using reactive bindings between inputs and outputs as well as  extensive pre-built widgets. The ability to run COPASI from R is provided by {\it CoRC}, the COPASI R Connector \cite{Forster2021}. This package provides an API to access to the functionality of COPASI from within the R language. Other packages used include {\it shinyTree} for the selection tree panel, {\it ggplot2} package to generate result plots, {\it DT} and  {\it formattable} to display and format tables. ShinyCOPASI uses the {\it xml} package to read the settings of optimization and parameter estimation tasks directly from the files.

The generic version of ShinyCOPASI is available at \href{http://shiny.copasi.org/}{http://shiny.copasi.org/}. The model-specific version is designed to be hosted and deployed in third party web servers; these will require to have installed the R system, a shiny server, as well as CoRC and the other dependencies. Detailed instructions for setting up a shiny server are available in the User Guide. Once the shiny server is set up, model-specific ShinyCOPASI interfaces can be deployed to allow access from web browsers. Multiple model-specific ShinyCOPASI instances can be deployed in the same shiny server.

\begin{figure}
	\centerline{\includegraphics[width=80mm]{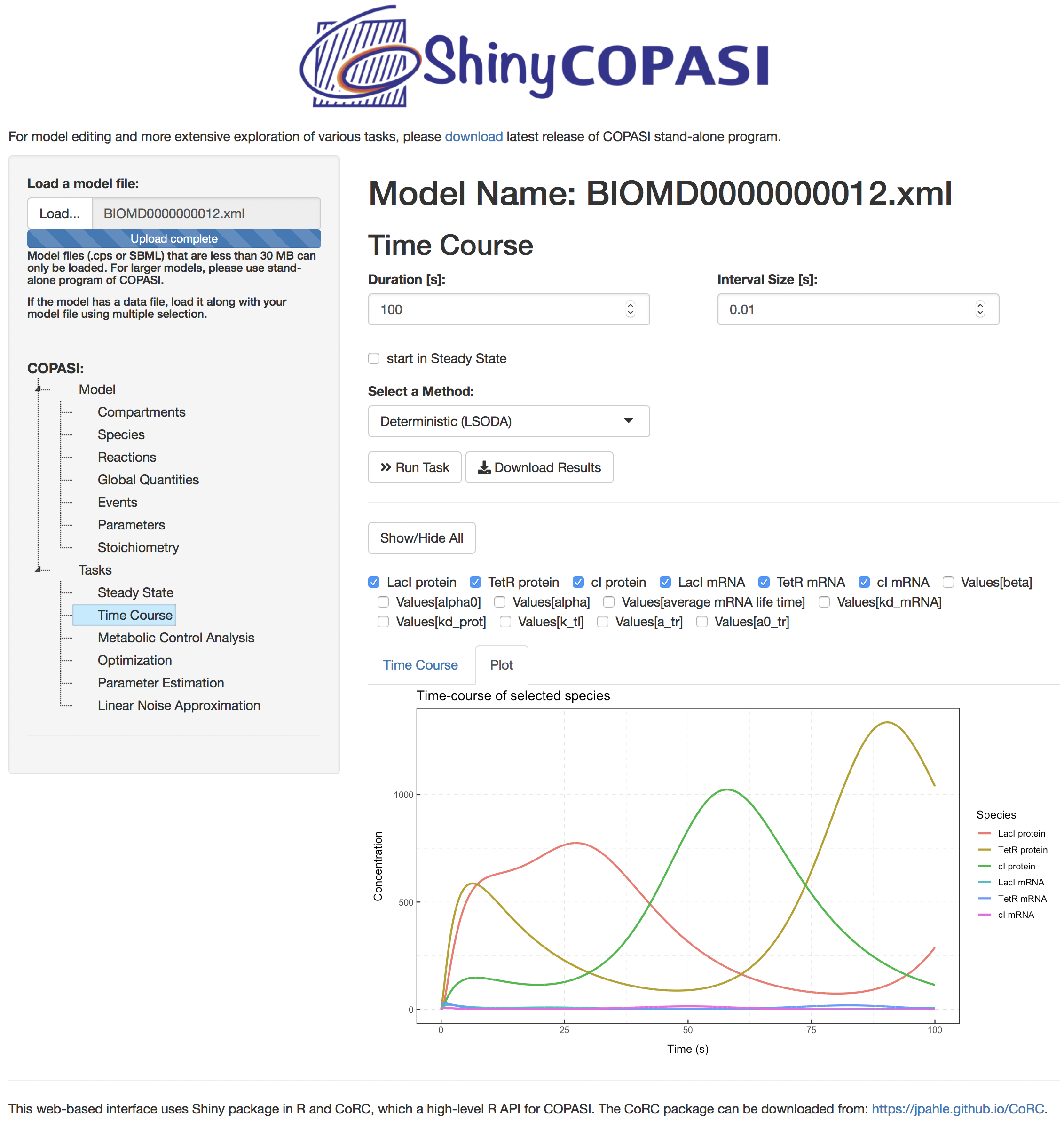}}
	\caption{ShinyCOPASI user interface showing results of a time course simulation for the repressilator model~\cite{Elowitz2000}, which was obtained from Biomodels in SBML format.}\label{fig:01}
\end{figure}

\section*{Interface}

ShinyCOPASI's interface (Fig.~{\ref{fig:01}}) mimics that of the COPASI standalone application as much as possible. It includes  three panels: one to choose model and tasks, another to display the task specifications, and another to display results. The generic and model-specific versions differ only in the model selection panel: the generic version allows the user to load a model (in COPASI .cps or SBML formats), while the model-specific version immediately loads the model hosted on the server. Additionally the latter also allows downloading the corresponding .cps file.

The leftmost panel allows the user to visualize the {\it Model} components, namely  {\it Compartments}, {\it Species}, {\it Reactions} {\it Parameters}, {\it Global Quantities} and {\it Events}. Note that this interface does not allow modifications to be made on the model. The {\it Tasks} section includes a subset of COPASI tasks, those currently supported by CoRC: {\it Steady State}, {\it Stoichiometry}, {\it Time Course}, {\it Metabolic Control Analysis}, {\it Optimization}, {\it Parameter Estimation} and {\it Linear Noise Approximation}. 

The top-right panel displays the model specifications and task settings, depending on the choice made in the {\it Selection} panel. Model details are displayed in a tabular form. For species, global quantities, and parameters an overview plot is also available which provides quick overview of magnitude differences between these items. When one task is selected, this panel displays its settings, a button to {\it run} the task and another to {\it download} its results (in .csv format). The bottom-right panel displays results of the corresponding task in a tabular form. In case of multiple results, these are displayed as multiple tables. For the {\it time course} task, there is an additional option to plot the results (see Fig.~\ref{fig:01}). 

To generate a model-specific version of ShinyCOPASI requires first loading the model in the regular standalone COPASI application and using a new option in the file menu: "Export Shiny Archive"  (Fig. 1 in User Guide). This generates a ZIP file containing all the files required for deployment in a shiny server installation, which will then provide access to that model alone. Thus, after installation of a ShinyCOPASI server it is easy to add new interfaces for specific models.

\section*{Discussion}
Our aim with ShinyCOPASI is to facilitate making COPASI (and SBML) models available on the web with a certain level of interactivity, including running some simulations and analyses. It is useful to allow quick inspection of a model online without requiring local software installation. This is, however, not intended to substitute the regular COPASI application, which is still required for large models, time intensive simulations and any kind of model editing.

The model-specific variant of ShinyCOPASI is particularly attractive to provide online access to published models. An example of this use case has been deployed in our group's research web site at \href{http://www.comp-sys-bio.org/models.html}{http://www.comp-sys-bio.org/models.html} where several of our published models are presented using this tool (follow links named ``run this model''). We hope that this will encourage model authors to provide web access to their models.

\section*{Acknowledgments}
We are grateful to the National Institutes of Health (NIGMS) for funding this work under Grant No. GM080219. We thank J Pahle and J Förster for CoRC.

\section*{Author Contributions}
A.G. and P.M. conceived and designed the experiments and wrote the paper; A.G.~performed the experiments and analyzed the data.

\section*{Conflict of interest}
The authors declare no financial conflict of interest. The founding sponsors had no role in the design of the study; in the collection, analyses, or interpretation of data; in the writing of the manuscript; or in the decision to publish the~results.

\nolinenumbers

%
%
%

\bibliography{references}
%
%
%
%

\end{document}